# Resolving the History of Life on Earth by Seeking Life *As We Know It* on Mars

**Authors:** Christopher E. Carr[1,2]*†.

**Affiliations:**

[1]Daniel Guggenheim School of Aerospace Engineering, Georgia Institute of Technology, Atlanta, GA, 30332, USA.

[2]School of Earth and Atmospheric Sciences, Georgia Institute of Technology, Atlanta, GA, 30332, USA.

*Correspondence to: cecarr@gatech.edu.

†Current address: ESM Building, Room G10, 620 Cherry St NW, Atlanta, GA 30332, USA.

**Abstract**

An origin of Earth life on Mars would resolve significant inconsistencies between the inferred history of life and Earth's geologic history. Life *as we know it* utilizes amino acids, nucleic acids, and lipids for the metabolic, informational, and compartment-forming subsystems of a cell. Such building blocks may have formed simultaneously from cyanosulfidic chemical precursors in a planetary surface scenario involving ultraviolet light, wet-dry cycling, and volcanism. However, early Earth was a water world, and the timing of the rise of oxygen on Earth is inconsistent with final fixation of the genetic code in response to oxidative stress. A cyanosulfidic origin of life could have taken place on Mars via photoredox chemistry, facilitated by orders of magnitude more sub-aerial crust than early Earth, and an earlier transition to oxidative conditions. Meteoritic bombardment may have generated transient habitable environments and ejected and transferred life to Earth. The Mars 2020 Perseverance Rover offers an unprecedented opportunity to confirm or refute evidence consistent with a cyanosulfidic origin of life on Mars, search for evidence of ancient life, and constrain the evolution of Mars' oxidation state over time. We should seek to prove or refute a Martian origin for life on Earth alongside other possibilities.

**One Sentence Summary:** A Martian origin for Earth life is consistent with the inferred history of life as we know it based on genomics and geology.

**Main Text:** Life *as we know it* utilizes amino acids, nucleic acids, and lipids for the metabolic, informational, and compartment-forming subsystems of a cell. Such building blocks may have formed simultaneously from cyanosulfidic chemical precursors in a planetary surface scenario involving ultraviolet light, wet-dry cycling, and volcanism (*1-3*). This process can be driven by photoredox chemistry with sulfite ($SO_3^{2-}$) mediating cycling of ferrocyanide ($[Fe^{II}(CN)_6]^{4-}$) and ferricyanide ($[Fe^{III}(CN)_6]^{3-}$) in combination with UV irradiation (*4*). While this scenario does not rule out other models such as an origin of life at seafloor vents (*5*), it plausibly and simultaneously addresses key challenges including formation and concentration of organic building blocks, their polymerization to yield functional molecules, and compartmentalization to yield proto-cellular entities (*6*).

All life as we know it shares a common ancestor (*7*). The most conserved genome regions occur within genes encoding the translation machinery (16S and 23S ribosomal subunits, transfer RNAs), which are themselves RNA machines involved in translating RNAs to polypeptides via the genetic code (*8*). These regions have changed little over 4 Ga (*9*). The deep evolutionary conservation of these molecular fossils is one piece of evidence for an RNA-Protein world preceding the DNA world (*10-13*). Furthermore, RNA molecules are capable of both storing hereditary information as well as catalyzing reactions, a dual role that may have been critical before the emergence of translation and the fixation of the genetic code. Protocell-like growth and division (*14*), for example, mediated by feedstock supply and/or photochemical processes (*15, 16*), could facilitate compartmentalization, selection, and evolution in the context of an RNA-Protein world.

Genetic evidence suggests that the last universal common ancestor (LUCA), which shares many features with modern life and was evolutionarily distant from its origin (*17*), inhabited an anoxic, "geochemically active environment rich in $H_2$, $CO_2$ and iron" (*18*). However, this setting on its own does not distinguish between sea floor vents and shallow-water hydrothermal habitats, nor between Earth and Mars.

Life as we know it utilizes dehydration synthesis to form the metabolic (protein, carbohydrate), informational (nucleic acid), and compartment-forming (lipid) polymers (*19*). Driving forces for dehydration include evaporation, sublimation, crystallization, or formation of hydrated minerals. Surface conditions thus offer plausible mechanisms to concentrate pre-biotic molecules and produce polymers. Dehydration could possibly occur due to nanoconfinement (*20*) in metal sulfides at alkaline vents, yet, at present, high water activity does not seem consistent with an origin of life (*21*).

**Early Earth was an Ocean World.** The cyanosulfidic origin theory is compelling, yet current data suggests that the early Earth was a water world (**Fig. 1A**), with little to no sub-aerial continental crust before 3.5 Ga (*22*), reaching 1-2% land by 3.0 Ga, and 5-8% by 2.5 Ga (*23, 24*). Consistent with these findings, recent analysis suggests initiation of continental weathering between 3 and 2.5 Ga (*25*). This would have limited the land area suitable for a cyanosulfidic origin of life to regions such as volcanic island hot spots (*22*).

**Land area on early Mars was orders of magnitude greater than early Earth.** Estimates of the water inventory on Mars at the Noachian/Hesperian boundary range from tens of meters (*26*) to ~1 km global equivalent layer (GEL) (*27*). The upper bound assumes a putative northern ocean after formation of the Martian dichotomy, e.g., the Borealis basin resulting from a massive impact event (*28*) at or before 4.2 Ga and possibly as old as 4.5 Ga (*29*). Even in the most extreme scenarios, Mars would have had orders of magnitude more land area than early Earth.

**The atmospheric rise of oxygen on Earth took place more than a billion years after fixation of the genetic code.** Additional support for a Mars origin of life on Earth comes from amino acids and the evolution of the genetic code itself. Granold et al. (*30*) proposed that the genetic code used a simpler set of amino acids and that the final diversification of amino acids happened in response to oxygen, suggesting that the diversification was late, e.g., coincident with the appearance of early local oxygen on Earth. It this hypothesis is correct, it would imply exposure to oxidizing conditions on the early Earth. However, the geologic record has revealed that Earth was largely devoid of oxygen for its first 2+ billion years, and appreciable quantities only accumulated after cyanobacteria invented oxygenic photosynthesis, resulting in the great oxidation event (GOE) (*31*).

While the GOE occurred around 2.33 Ga (*32*), evidence for early local oxidative weathering suggests there could have been transient local oxygen pulses at or before 3 Ga (*33,*

*34*). Nevertheless, the late diversification is problematic because the root of the archaeal phylum Euryarchaeota, which arose after LUCA, has been dated to >4 Ga (*35*) (**Fig. 1D**), and the genetic code must have been fully established before this time (*18, 36*) including use of selenocysteine by LUCA (*18*). Time-calibrated phylogenomics extending this deep into life's history comes with caveats and wide error bars, yet it highlights a >1.5 Ga inconsistency. Thus, the timing of the fixation of the genetic code does not align with the inferred oxidation state of the early Earth. However, on Mars, surface conditions became oxidizing much earlier.

**A Cyanosulfidic Origin of Life on Mars?** Today Mars is dry and cold (*37*), yet early Mars was habitable for life *as we know it*, with significantly more water (*27, 38, 39*); availability of the main elements used by life (C, H, N, O, P, and S) and energy sources including variable redox states of iron and sulfur minerals are recorded in the stratigraphy of Gale Crater (*40, 41*). Widespread clay minerals confirm extensive periods of subsurface water-rock interactions before 3.7 Ga (**Fig. 1C**), yet also suggest even early Mars had mostly cold and relatively arid surface conditions (*39*), which could have aided accumulation of organic molecules through concentration and low hydrolysis rates.

Sasselov et al. (*3*) delineate a plausible pathway for a cyanosulfidic origin of life on early Mars. They suggest that igneous intrusions, volcanism, or impacts interacting with cyanide salt deposits could have generated the relevant feedstocks to produce nucleotide, amino acid, and lipid precursors. Here, we also suggest a Mars cyanosulfidic origin of life could have seeded life on Earth, resolving the inconsistencies previously noted for an Earth origin of life on Earth.

Lightning would have provided hydrogen cyanide (HCN), representing a total fixed nitrogen budget on par with that of the early Earth (*42*). Meteorite impacts (**Fig. 1C**) may also have generated HCN and provided phosphate (*43*). Sulfite would likely have been available on early Mars as a consequence of volcanic $SO_2$, a $CO_2$ atmosphere, and low temperatures (*44*); later oxidation would have led to the formation of sulfate minerals, consistent with remote sensing (*45*) and in situ measurements (*46*). The presence of sulfites would also help to explain the relative dearth of carbonates on Mars (*44*).

**Could the RNA world have originated on Mars?** A cyanosulfidic origin of life (*2, 3*) would produce all the building blocks required for an RNA-protein world (*11, 13*). What we are still lacking is knowledge about whether ancient Mars, especially before the Noachian-Hesperian boundary (**Fig. 1**), was conducive to the formation, stabilization, and evolution of an RNA-Protein world and ultimately cellular life.

Hints are beginning to emerge: Mojarro et al. (*47*) evaluated plausible divalent cations ($Fe^{2+}$, $Mg^{2+}$, and $Mn^{2+}$) at different pH conditions and found the lowest rate of metal-catalyzed hydrolysis of RNA for $Mg^{2+}$ at pH 5.4, rather more acidic than the circumneutral conditions suggested by the cyanosulfidic model (*3*). Bray et al. (*9*) found that $Fe^{2+}$ and $Mn^{2+}$ can replace the modern role of $Mg^{2+}$ for ribosomal RNA folding and translation in anoxic conditions, relevant to early Earth or early Mars. An improved understanding of Mars redox history is necessary to understand the potential origin and evolution of an RNA world on Mars.

For all known life, DNA ultimately replaced RNA as the carrier of hereditary information. However, if an RNA-protein world evolved on Mars, we may still be able to sample fossilized or extant traces of it in the subsurface, or may find, like all known Earth life, that its role in heredity has been displaced by DNA.

**Mars surface redox evolution occurred earlier than on Earth.** While an oxidizing atmosphere on Earth arose due to biological production of oxygen, oxidation on Mars is thought to have resulted from crustal water sequestration (*48, 49*) as well as photolysis of water and subsequent loss of hydrogen (*27, 38, 50, 51*). The present-day surface of Mars is highly oxidizing, and this state has likely persisted for billions of years as inferred from manganese hydroxides identified at Gale and Endeavor craters (*52*).

On Mars, the transition to more oxidizing conditions is recorded by the deposition of sulfates ($SO_4^{2-}$ with oxidation state $S^{6+}$) from approximately the Noachian-Hesperian transition (3.7 Ga) onward (*46, 53*) (**Fig. 1C**). This is consistent with orbital observations of Noachian-aged $Fe^{3+}$-bearing clay minerals (e.g., associated with surface oxidation) and in situ detection of their purported ferrous smectite precursors (*54*). As the full range of sulfate deposits on Mars has yet to be identified, oxidation may also have occurred earlier.

This timing is consistent with a potential role of oxidation in shaping fixation of the genetic code, followed by meteoritic transfer of life to Earth. This resolves the problematic interpretation (*30*) that the genetic code was not fixed until 1.5 Ga after the timing predicted by genetic evidence (*18, 35, 36*) and is consistent with the antiquity of antioxidant enzymes and recent work suggesting a potential role for reactive sulfur species in their evolution (*55*).

**Phosphate availability was likely higher on early Mars than early Earth.** Early nucleic-acid based life would have required phosphate to build informational polymers and store energy. Release rates for phosphate-bearing minerals are estimated to be 45 times higher for Mars compared to Earth, with ~2x higher equilibrium phosphate concentrations (*56*). In contrast, phosphorus availability was limited on the early Earth due to a dearth of oxidizing power that limited recycling until the great oxidation event (GOE) (*57*). Cyanide can promote release of phosphate from iron phosphates, enabling generation of organophosphates under desiccating conditions (*58*); dry conditions were widespread on early Mars but lacking on early Earth.

**Ejecta from large impact events could have transferred viable life from Mars to Earth.** The plausibility of lithopanspermia has been theoretically and experimentally tested, including non-sterilizing ejection from planetary surfaces as measured in Mars meteorite ALH84001 (<40°C based on residual magnetization) (*59*) and through experimental studies of microbial survival to hypervelocity shock pressures (*60*). Modeling suggests that the mass of Mars to Earth viable transfers (*61, 62*) is greater than $10^{12}$ kg (*62*), and this ignores potential transfers before 3.5 Ga when cratering rates were higher (**Fig. 1C**). Recent data supporting a Martian dynamo at 4.5 and 3.7 Ga (*29*) suggests any life being ejected before 3.7 Ga would have been afforded protection from space radiation while on Mars, although perhaps not for a few hundred million years (3.4 to 3.45 Ga) after its arrival on Earth (*63*) (**Fig. 1B**).

Organisms arriving on Earth—putative chemolithoautotrophs—would have adapted to available redox gradients, including hydrothermal systems. While thermophily near the base of the tree of life was originally interpreted to provide evidence of the vent origin model, it can also be explained by parallel adaptation to high temperatures (*64*), such as selection for heat resistant species during Mars-Earth meteoritic transfer, or via subsequent heating of Earth's oceans during impact events. Reconstructed optimal growth temperature (OGT) across the tree of life suggest hyperthermophily at the base of the Archaea and Bacteria domains, but a colder OGT for LUCA of ~45°C (confidence interval 33-68°C) (*65*), consistent with non-sterilizing temperatures during meteoritic ejection (*59*).

**A Mars Origin for Life on Earth:** What we know about early Mars is consistent with the cyanosulfidic origin of life, and conditions there would have provided orders of magnitude more land area for prebiotic chemistry to cross the threshold to life. The Noachian-Hesperian transition and commensurate changes in surface redox balance on Mars are also consistent with the timing of fixation of the genetic code (*12, 36*) and the antiquity of antioxidant enzymes (*55*). Theoretical (*61, 62*) and experimental (*59, 60*) studies of lithopanspermia imply that Earth, Mars, and Venus would have shared viable microbes (*62*), if present in ejecta, and subsequent genetic evidence is consistent with this timeline.

Other options are possible (*66*): If cyanosulfidic chemistry and the transition to life is a planetary phenomenon (*3*), relatively fast and easy given the right conditions, then we may find that such life evolved independently on both early Mars and early Earth, despite the latter's dearth of sub-aerial crust. If booting up life is harder or rare, this would raise the prior that we may share common ancestry with any life on Mars. Finding no life on Mars would be incredibly revealing, because it would suggest life is a rare event requiring very particular circumstances.

Given the inconsistencies between the inferred history of life as we know it and Earth's geologic history, it is plausible that life on Earth could have originated on Mars (*66*). Such a history, while incredible, is a story of our past that explains the available genetic and geological evidence. It is also testable.

**Mars 2020 and Future Missions Should Search for Life As We Know It.** While the Perseverance Rover will focus on habitability and evidence of ancient, and not extant, life (*67*), it can seek to confirm or refute conditions consistent with a cyanosulfidic origin of life (*3*), as well as seek chemical precursors to life and identify high abundance organics.

Jezero Crater is a 45 km diameter impact crater that once harbored an open-basin lake system recorded by sedimentary deposits, deltas, and other features associated with fluvial activity that may have ended by around the Noachian-Hesperian boundary or earlier (*68*). Analysis of the western fan deposit at the crater inlet has identified clay minerals (Fe/Mg-smectite) and stratigraphy that likely records an extended history of early Mars, drawn from a watershed much larger than the crater, which includes smectites, olivine, and Mg-rich carbonate-bearing terrains, among others (*69*). Beyond likely having been habitable itself, the Jezero Lake and associated watershed should provide access to samples originating in a diverse set of other potentially habitable environments, and the lacustrine setting implies access to a well-ordered stratigraphic record of the Noachian period on Mars.

Because the fluvial activity associated with lacustrine sedimentary deposits at Jezero Crater likely ended by 3.8 Ga, evidence within those deposits of increasing oxidation over time would be consistent with the ancestral relationship scenario. Alternatively, a relatively reducing surface environment up to 3.8 Ga would not rule out life as we know it on Mars, but would weaken the case for any ancestral relationship, at least one linked to the Jezero Crater source region. For comparison, sedimentary rocks within Gale Crater, which formed between 3.6 and 3.8 Ga, record a mix of oxidation states even within the base of the stratigraphic section (*70*), as well as evidence of redox stratification in younger deposits (*40*), consistent with an oxidative surface environment by the Noachian-Hesperian boundary or early Hesperian.

If putative ancient life on Mars was exposed to an oxidizing surface environment before the earliest evidence of life on Earth, a Mars-Earth transfer is not ruled out. If no part of Mars became oxidizing until significantly after the earliest signs of life on Earth (< 3.7 Ga), or certainly after the earliest unambiguous evidence of fossil life on Earth (3.5 Ga) (*71, 72*), then the window of opportunity to complete fixation of the genetic code on an oxidized Mars prior to a hypothesized Mars-Earth transit would close, and lead to rejection of the hypothesis that oxidizing conditions on Mars could explain fixation of the genetic code.

Silica has high preservation potential and hydrated silica deposits have been identified within Jezero crater in association with delta deposits (*73*). Organic materials can be concentrated locally and stabilized by adsorbing to silica and the aforementioned clays (Fe/Mg-smectite). Lacustrine carbonates in Jezero Crater also offer high preservation potential (*74*). A study of amino acid preservation in simulated Mars conditions found the highest preservation in smectites and sulfates (*75*).

Extrapolation from modeling of hydrolysis and ancient DNA sequencing projects (*76, 77*) suggests that 100 bp DNA at –60˚C could have a half-life >2 Ga. However, other factors such as naturally-occurring radioisotope decay may limit the decay half-life to perhaps 10 My (*78*). RNA generally has higher hydrolysis rates, although the pH dependence is different (*47*). Thus, direct detection of nucleic acid polymers in Noachian-aged terrain is extremely improbable. However, nucleobases, the informational component of nucleic acids, have been identified in meteorites (*79*), and thus, at least under cold conditions, nucleobases can survive >4 Ga.

Intact nucleic acids and nucleobases can adsorb to silica, and in the relative absence of later aqueous and thermal alterations, nucleobases can be retained. Silica is, in fact, used to isolate cross-linked RNA and proteins to study their interactions (*80*). Cross-linking would be expected as part of the diagenesis process, as would be deamination of cytosine to yield uracil (*81*). Mars 2020's instruments are not designed to specifically interrogate uracil, but it could be targeted in returned samples along with other nucleobases.

Lipids are also one of the most geologically stable biomolecules, and are used as a biosignature in studies of Earth life in rocks older than 2 Ga (*82*). Studies of Mars-relevant diagenesis reveal caveats. Tan et al. (*83*) found that hydrous pyrolysis from 200-280˚C for 72 hours significantly degraded lipids, especially with low carbon content, high water-to-rock ratios, and iron-rich minerals (*83*). Carrizo et al. (*84*) demonstrated the potential to detect lipid biomarkers using a tunable Raman laser spectrometer and found that of silica rich vs. iron rich samples tested, the silica rich material enhanced preservation. Thus, silica-rich deposits in Jezero Crater could be targeted in the search for nucleobases and lipids.

Many of the mineral deposits mentioned are expected to be accessible to the Mars 2020 Rover, for example, within the stratigraphy of the western delta deposits. Because Mars lacks a substantial atmosphere, organics within deposits exposed at the surface would be damaged over time by space radiation. To avoid this damage, low exposure ages are desirable. Mudstone on the floor of Gale Crater was found to have a exposure age of 78 ± 30 million years (*85*) highlighting the potential to utilize wind-driven geomorphological change to obtain samples with potential for complex organic molecules as undegraded by cosmic rays as possible. Thus, it may be feasible to use a similar approach in Jezero Crater, leveraging prior wind analysis (*86*) and *in situ* measurements (MEDA) to assess access to low exposure age stratigraphy. Because accessible exposure ages may still be large, drilling, including, in future missions, below the 1-2 meter penetration depth of space radiation, remains critical for accessing undegraded samples.

While the capabilities of the Mars 2020 Rover are unprecedented, and several instruments (SuperCam (*87*), SHERLOC (*88*), PIXL (*89*)) will be used to detect or infer the presence of organics and specific chemical moieties, they may lack the ability to uniquely identify specific molecules. Instruments in development could one day target poly-peptides and nucleic acids (*90, 91*).

Mars 2020 is uniquely positioned to seek evidence consistent with a cyanosulfidic origin of life on Mars, search for ancient evidence of life, and constrain the evolution of Mars' oxidation state over time. Unlike Earth, where ancient rocks are rare, on Mars we can access rocks across nearly the full 4.5 Ga history of that planet (*92*). Despite the modern thin (1% that of Earth) atmosphere and cold average surface temperature (–60˚C), the subsurface of Mars likely remains habitable (*93*). Future missions, with access to special regions, including deep drilling to seek habitable environments in the sub-surface, may be required to target any extant life and unambiguously determine whether life as we know it exists on Mars today and if so, whether it is related to us.

**References and Notes:**

1. B. Damer, D. Deamer, The Hot Spring Hypothesis for an Origin of Life. *Astrobiology*, ast.2019.2045 (2019).
2. B. H. Patel, C. Percivalle, D. J. Ritson, C. D. Duffy, J. D. Sutherland, Common origins of RNA, protein and lipid precursors in a cyanosulfidic protometabolism. *Nat Chem* **7**, 301-307 (2015).
3. D. D. Sasselov, J. P. Grotzinger, J. D. Sutherland, The origin of life as a planetary phenomenon. *Science Advances* **6**, eaax3419 (2020).
4. J. Xu *et al.*, Photochemical reductive homologation of hydrogen cyanide using sulfite and ferrocyanide. *Chem. Commun. (Camb.)* **54**, 5566-5569 (2018).
5. W. Martin, J. Baross, D. Kelley, M. J. Russell, Hydrothermal vents and the origin of life. *Nat Rev Microbiol* **6**, 805-814 (2008).
6. G. F. Joyce, J. W. Szostak, Protocells and RNA Self-Replication. *Cold Spring Harbor Perspectives in Biology* **10**, a034801 (2018).
7. N. R. Pace, The universal nature of biochemistry. *Proc Natl Acad Sci USA* **98**, 805-808 (2001).
8. T. A. Isenbarger *et al.*, The most conserved genome segments for life detection on Earth and other planets. *Origins of Life and Evolution of Biospheres* **38**, 517-533 (2008).
9. M. S. Bray *et al.*, Multiple prebiotic metals mediate translation. *Proc Natl Acad Sci USA* **115**, 12164 (2018).
10. C. Hsiao, S. Mohan, B. Kalahar, L. Williams, Peeling the Onion: Ribosomes Are Ancient Molecular Fossils. *Mol Biol Evol* **26**, 2415-2425 (2009).
11. A. D. Goldman, R. Samudrala, J. A. Baross, The evolution and functional repertoire of translation proteins following the origin of life. *Biol Direct* **5**, 15-12 (2010).
12. G. Fournier, J. Neumann, J. Gogarten, Inferring the Ancient History of the Translation Machinery and Genetic Code via Recapitulation of Ribosomal Subunit Assembly Orders. *PLoS ONE* **5**, e9437 (2010).
13. A. Harish, G. Caetano-Anollés, Ribosomal history reveals origins of modern protein synthesis. *PLoS ONE* **7**, e32776 (2012).
14. I. Budin, J. W. Szostak, Physical effects underlying the transition from primitive to modern cell membranes. *Proc Natl Acad Sci USA* **108**, 5249-5254 (2011).

15. T. F. Zhu, K. Adamala, N. Zhang, J. W. Szostak, Photochemically driven redox chemistry induces protocell membrane pearling and division. *Proc Natl Acad Sci USA* **109**, 9828-9832 (2012).
16. J. W. Szostak, The Narrow Road to the Deep Past: In Search of the Chemistry of the Origin of Life. *Angew Chem Int Ed Engl* **56**, 11037-11043 (2017).
17. M. D. Cantine, G. P. Fournier, Environmental Adaptation from the Origin of Life to the Last Universal Common Ancestor. *Origins of Life and Evolution of Biospheres* **48**, 35-54 (2018).
18. M. C. Weiss *et al.*, The physiology and habitat of the last universal common ancestor. *Nature Microbiology* **1**, 1-8 (2016).
19. N. Kitadai, S. Maruyama, Origins of building blocks of life: A review. *Geoscience Frontiers* **9**, 1117-1153 (2018).
20. D. Muñoz-Santiburcio, D. Marx, Chemistry in nanoconfined water. *Chem Sci* **8**, 3444-3452 (2017).
21. M. Frenkel-Pinter, V. Rajaei, J. B. Glass, N. V. Hud, L. D. Williams, Water and Life: The Medium is the Message. *Journal of Molecular Evolution*,  (2021).
22. J. L. Bada, J. Korenaga, Exposed Areas Above Sea Level on Earth >3.5 Gyr Ago: Implications for Prebiotic and Primitive Biotic Chemistry. *Life* **8**, 55 (2018).
23. S. V. Lalonde, K. O. Konhauser, Benthic perspective on Earth's oldest evidence for oxygenic photosynthesis. *Proceedings of the National Academy of Sciences*,  (2015).
24. N. Flament, N. Coltice, P. F. Rey, The evolution of the 87Sr/86Sr of marine carbonates does not constrain continental growth. *Precambrian Research* **229**, 177-188 (2013).
25. B. W. Johnson, B. A. Wing, Limited Archaean continental emergence reflected in an early Archaean 18 O-enriched ocean. *Nature Geosci* **13**, 243-248 (2020).
26. M. H. Carr, J. W. Head, Martian surface/near-surface water inventory: Sources, sinks, and changes with time. *Geophysical Research Letters* **42**, 726-732 (2015).
27. H. Kurokawa *et al.*, Evolution of water reservoirs on Mars: Constraints from hydrogen isotopes in martian meteorites. *Earth and Planetary Science Letters* **394**, 179-185 (2014).
28. J. C. Andrews-Hanna, M. T. Zuber, W. B. Banerdt, The Borealis basin and the origin of the martian crustal dichotomy. *Nature* **453**, 1212-1215 (2008).
29. A. Mittelholz, C. L. Johnson, J. M. Feinberg, B. Langlais, R. J. Phillips, Timing of the martian dynamo: New constraints for a core field 4.5 and 3.7 Ga ago. *Science Advances* **6**, eaba0513 (2020).
30. M. Granold, P. Hajieva, M. I. Toşa, F.-D. Irimie, B. Moosmann, Modern diversification of the amino acid repertoire driven by oxygen. *Proc Natl Acad Sci USA* **115**, 41-46 (2018).
31. A. Harel, S. Karkar, S. Cheng, P. G. Falkowski, D. Bhattacharya, Deciphering Primordial Cyanobacterial Genome Functions from Protein Network Analysis. *Current Biology* **25**, 628-634 (2015).
32. G. Luo *et al.*, Rapid oxygenation of Earth's atmosphere 2.33 billion years ago. *Science Advances* **2**, e1600134 (2016).
33. J. Mukhopadhyay *et al.*, Oxygenation of the Archean atmosphere: New paleosol constraints from eastern India. *Geology* **42**, 923-926 (2014).
34. S. A. Crowe *et al.*, Atmospheric oxygenation three billion years ago. *Nature* **501**, 535-538 (2013).

35. J. M. Wolfe, G. P. Fournier, Horizontal gene transfer constrains the timing of methanogen evolution. *Nat Ecol Evol* **2**, 897-903 (2018).
36. G. P. Fournier, E. J. Alm, Ancestral Reconstruction of a Pre-LUCA Aminoacyl-tRNA Synthetase Ancestor Supports the Late Addition of Trp to the Genetic Code. *Journal of Molecular Evolution* **80**, 171-185 (2015).
37. H. Savijärvi, T. H. McConnochie, A.-M. Harri, M. Paton, Water vapor mixing ratios and air temperatures for three martian years from Curiosity. *Icarus* **326**, 170-175 (2019).
38. N. R. Alsaeed, B. M. Jakosky, Mars Water and D/H Evolution From 3.3 Ga to Present. *J. Geophys. Res. Planets* **124**, 3344-3353 (2019).
39. B. L. Ehlmann *et al.*, Subsurface water and clay mineral formation during the early history of Mars. *Nature* **479**, 53-60 (2011).
40. J. A. Hurowitz *et al.*, Redox stratification of an ancient lake in Gale crater, Mars. *Science* **356**, eaah6849 (2017).
41. J. P. Grotzinger *et al.*, Deposition, exhumation, and paleoclimate of an ancient lake deposit, Gale crater, Mars. *Science* **350**, aac7575-aac7575 (2015).
42. A. Segura, Nitrogen fixation on early Mars by volcanic lightning and other sources. *Geophys. Res. Lett.* **32**, 173 (2005).
43. M. A. Pasek, D. S. Lauretta, Aqueous corrosion of phosphide minerals from iron meteorites: a highly reactive source of prebiotic phosphorus on the surface of the early Earth. *Astrobiology* **5**, 515-535 (2005).
44. G. M. Marion, J. S. Kargel, J. K. Crowley, D. C. Catling, Sulfite–sulfide–sulfate–carbonate equilibria with applications to Mars. *Icarus* **225**, 342-351 (2013).
45. J. Flahaut, C. Quantin, P. Allemand, P. Thomas, L. Le Deit, Identification, distribution and possible origins of sulfates in Capri Chasma (Mars), inferred from CRISM data. *J. Geophys. Res.* **115**, 499 (2010).
46. W. Rapin *et al.*, An interval of high salinity in ancient Gale crater lake on Mars. *Nature Geosci* **12**, 889-895 (2019).
47. A. Mojarro, L. Jin, J. W. Szostak, J. W. Head, M. T. Zuber, In search of the RNA world on Mars. *bioRxiv* **6**, 2020.2002.2028.964486 (2020).
48. J. Tuff, J. Wade, B. J. Wood, Volcanism on Mars controlled by early oxidation of the upper mantle. *Nature* **498**, 342-345 (2013).
49. J. Wade, B. Dyck, R. M. Palin, J. D. P. Moore, A. J. Smye, The divergent fates of primitive hydrospheric water on Earth and Mars. *Nature* **552**, 391-394 (2017).
50. H. Hartman, C. P. McKay, Oxygenic photosynthesis and the oxidation state of Mars. *Planetary and Space Science* **43**, 123-128 (1995).
51. A. A. Fedorova *et al.*, Stormy water on Mars: The distribution and saturation of atmospheric water during the dusty season. *Science* **367**, 297 (2020).
52. N. Noda *et al.*, Highly Oxidizing Aqueous Environments on Early Mars Inferred From Scavenging Pattern of Trace Metals on Manganese Oxides. *J. Geophys. Res. Planets* **75**, 1-1295 (2019).
53. K. Fukushi, Y. Sekine, H. Sakuma, K. Morida, R. Wordsworth, Semiarid climate and hyposaline lake on early Mars inferred from reconstructed water chemistry at Gale. *Nat Comms* **10**, 1-11 (2019).
54. S. M. Chemtob, R. D. Nickerson, R. V. Morris, D. G. Agresti, J. G. Catalano, Oxidative Alteration of Ferrous Smectites and Implications for the Redox Evolution of Early Mars. *J. Geophys. Res. Planets* **122**, 2469-2488 (2017).

55. A. Neubeck, F. Freund, Sulfur Chemistry May Have Paved the Way for Evolution of Antioxidants. *Astrobiology* **20**, 670-675 (2020).
56. C. T. Adcock, E. M. Hausrath, P. M. Forster, Readily available phosphate from minerals in early aqueous environments on Mars. *Nature Geosci* **6**, 824-827 (2013).
57. M. A. Kipp, E. E. Stueken, Biomass recycling and Earth's early phosphorus cycle. *Sci Adv* **3**, eaao4795 (2017).
58. B. Burcar *et al.*, A Stark Contrast to Modern Earth: Phosphate Mineral Transformation and Nucleoside Phosphorylation in an Iron- and Cyanide-Rich Early Earth Scenario. *Angewandte Chemie International Edition* **58**, 16981-16987 (2019).
59. B. P. Weiss *et al.*, A low temperature transfer of ALH84001 from Mars to Earth. *Science* **290**, 791-795 (2000).
60. G. Horneck *et al.*, Microbial rock inhabitants survive hypervelocity impacts on Mars-like host planets: first phase of lithopanspermia experimentally tested. *Astrobiology* **8**, 17-44 (2008).
61. C. Mileikowsky *et al.*, Natural transfer of viable microbes in space. *Icarus* **145**, 391-427 (2000).
62. R. J. Worth, S. Sigurdsson, C. H. House, Seeding life on the moons of the outer planets via lithopanspermia. *Astrobiology* **13**, 1155-1165 (2013).
63. J. A. Tarduno *et al.*, Geodynamo, Solar Wind, and Magnetopause 3.4 to 3.45 Billion Years Ago. *Science* **327**, 1238 (2010).
64. B. Boussau, S. Blanquart, A. Necsulea, N. Lartillot, M. Gouy, Parallel adaptations to high temperatures in the Archaean eon. *Nature* **456**, 942-945 (2008).
65. M. Groussin, B. Boussau, S. Charles, S. Blanquart, M. Gouy, The molecular signal for the adaptation to cold temperature during early life on Earth. *Biology Letters* **9**, 20130608 (2013).
66. C. P. McKay, An Origin of Life on Mars. *Cold Spring Harbor Perspectives in Biology* **2**, a003509-a003509 (2010).
67. K. H. Williford *et al.*, in *From Habitability to Life on Mars*. (Elsevier, 2018), pp. 275-308.
68. T. A. Goudge, D. Mohrig, B. T. Cardenas, C. M. Hughes, C. I. Fassett, Stratigraphy and paleohydrology of delta channel deposits, Jezero crater, Mars. *Icarus* **301**, 58-75 (2018).
69. T. A. Goudge, J. F. Mustard, J. W. Head, C. I. Fassett, S. M. Wiseman, Assessing the mineralogy of the watershed and fan deposits of the Jezero crater paleolake system, Mars. *J. Geophys. Res. Planets* **120**, 775-808 (2015).
70. J. P. Grotzinger *et al.*, A habitable fluvio-lacustrine environment at Yellowknife Bay, Gale crater, Mars. *Science* **343**, 1242777-1242777 (2014).
71. J. W. Schopf, Microfossils of the Early Archean Apex Chert: New Evidence of the Antiquity of Life. *Science* **260**, 640 (1993).
72. J. W. Schopf, K. Kitajima, M. J. Spicuzza, A. B. Kudryavtsev, J. W. Valley, SIMS analyses of the oldest known assemblage of microfossils document their taxon-correlated carbon isotope compositions. *Proceedings of the National Academy of Sciences of the United States of America* **115**, 53-58 (2018).
73. J. D. Tarnas *et al.*, Orbital Identification of Hydrated Silica in Jezero Crater, Mars. *Geophys. Res. Lett.* **46**, 12771-12782 (2019).

74. B. H. N. Horgan, R. B. Anderson, G. Dromart, E. S. Amador, M. S. Rice, The mineral diversity of Jezero crater: Evidence for possible lacustrine carbonates on Mars. *Icarus* **339**, 113526 (2020).
75. R. dos Santos, M. Patel, J. Cuadros, Z. Martins, Influence of mineralogy on the preservation of amino acids under simulated Mars conditions. *Icarus* **277**, 342-353 (2016).
76. C. D. Millar, D. M. Lambert, Ancient DNA: Towards a million-year-old genome. *Nature* **499**, 34-35 (2013).
77. M. E. Allentoft *et al.*, The half-life of DNA in bone: measuring decay kinetics in 158 dated fossils. *Proceedings Biological sciences / The Royal Society* **279**, 4724-4733 (2012).
78. G. Kminek, J. L. Bada, K. Pogliano, J. F. Ward, Radiation-dependent limit for the viability of bacterial spores in halite fluid inclusions and on Mars. *Radiation research* **159**, 722-729 (2003).
79. M. P. Callahan *et al.*, Carbonaceous meteorites contain a wide range of extraterrestrial nucleobases. *Proceedings of the National Academy of Sciences of the United States of America* **108**, 13995-13998 (2011).
80. C. Asencio, A. Chatterjee, M. W. Hentze, Silica-based solid-phase extraction of cross-linked nucleic acid–bound proteins. *Life Sci Alliance* **1**, e201800088-e201800088 (2018).
81. J. Dabney, M. Meyer, S. Pääbo, Ancient DNA Damage. *Cold Spring Harb Perspect Biol* **5**, a012567-a012567 (2013).
82. K. L. French *et al.*, Reappraisal of hydrocarbon biomarkers in Archean rocks. *Proc Natl Acad Sci USA* **112**, 5915-5920 (2015).
83. J. Tan, M. A. Sephton, Organic Records of Early Life on Mars: The Role of Iron, Burial, and Kinetics on Preservation. *Astrobiology* **20**, 53-72 (2020).
84. D. Carrizo *et al.*, Detection of Potential Lipid Biomarkers in Oxidative Environments by Raman Spectroscopy and Implications for the ExoMars 2020-Raman Laser Spectrometer Instrument Performance. *Astrobiology* **20**, 405-414 (2020).
85. K. A. Farley *et al.*, In Situ Radiometric and Exposure Age Dating of the Martian Surface. *Science* **343**, 1247166 (2014).
86. M. Day, T. Dorn, Wind in Jezero Crater, Mars. *Geophys. Res. Lett.* **46**, 3099-3107 (2019).
87. R. C. Wiens, S. Maurice, F. R. Perez, The SuperCam remote sensing instrument suite for the Mars 2020 rover mission: A preview. *Spectroscopy* **32**, 50-55 (2017).
88. L. Beegle *et al.*, paper presented at the IEEE Aerospace Conference, 2013.
89. A. Allwood *et al.*, paper presented at the IEEE Aerospace Conference, 2015.
90. A. Mojarro *et al.*, Nucleic Acid Extraction and Sequencing from Low-Biomass Synthetic Mars Analog Soils for In Situ Life Detection. *Astrobiology* **19**, 1139-1152 (2019).
91. C. E. Carr *et al.*, paper presented at the IEEE Aerospace Conference, 2016.
92. L. C. Bouvier *et al.*, Evidence for extremely rapid magma ocean crystallization and crust formation on Mars. *Nature* **558**, 586-589 (2018).
93. E. G. Jones, C. H. Lineweaver, J. D. Clarke, An extensive phase space for the potential martian biosphere. *Astrobiology* **11**, 1017-1033 (2011).

**Acknowledgments:** I thank Maria T. Zuber and Gary Ruvkun for extensive discussions and collaborative research on this topic over the last 15 years. **Funding:** C.E.C. was supported by

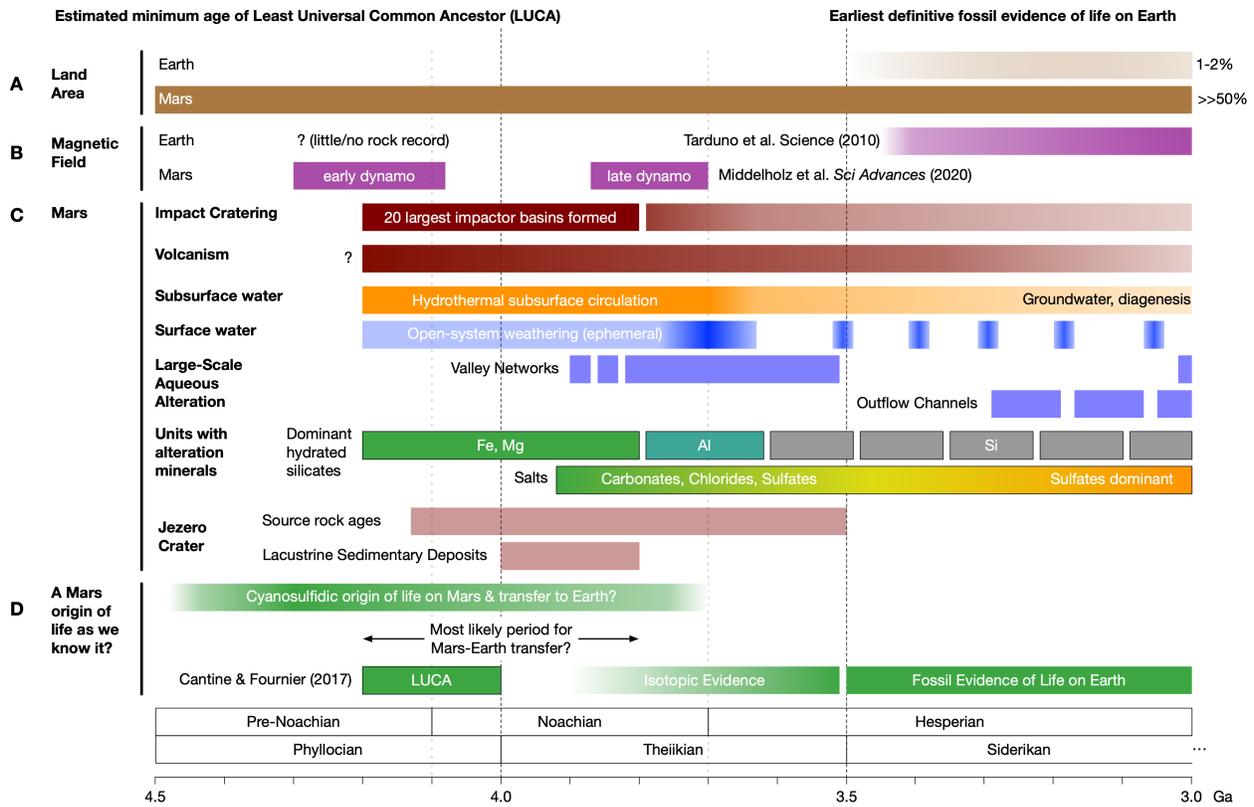

**Fig. 1. Planetary context for a hypothesized cyanosulfidic origin of life on Mars and its transport to Earth via lithopanspermia.** The horizontal axis for all panels is billions of years ago (Ga). (A) Early Mars had orders of magnitude more land area available to support a cyanosulfidic origin of life. (B) Life arising on early Mars would have been protected by a magnetic field. The lack of a rock record prevents similar knowledge of the early Earth, although it is likely Earth's dynamo had started by 3.4 to 3.45 Ga (*63*). (C) Impact cratering on early Mars, especially from ~4.2 to ~3.8 Ga, would have provided both habitable environments for life to arise as well as facilitate its transport to Earth via meteoritic ejecta. Volcanism would have played a role in facilitating a cyanosulfidic origin of life (*3*) in combination with hydrothermal subsurface activity and ephemeral surface waters. The aqueous record of this time and the geochemical transformation of Mars is recorded by the presence of hydrated silicates and salts, with sulfates as a particularly important record of early oxidation on Mars in comparison to Earth, where widespread oxidation occurred over a billion years later. (D) If a cyanosulfidic origin of life occurred on early Mars, such life could have been transferred to Earth around the time of, but not necessarily coincident with, the Last Universal Common Ancestor (LUCA). For example, LUCA may have existed on Mars and life on Earth could have arrived via multiple transfer events. Estimated timing of Mars processes are as reported by Ehlmann et al. (*39*).